# MQG4AI: Towards Responsible High-risk AI – Illustrated for Transparency Focusing on Explainability Techniques

Miriam Elia[a,*], Alba Maria Lopez[b], Katherin Alexandra Corredor Páez[b], Bernhard Bauer[a] and Esteban Garcia-Cuesta[b,*]

[a]*Faculty of Applied Computer Science, University of Augsburg, Universitätsstraße 6a, Augsburg, 86159, Bavaria, Germany*
[b]*Artificial Intelligence Department, ETSI Informaticos, Universidad Politécnica, C. de los Ciruelos, 28660 Madrid, Spain*

### ABSTRACT

As artificial intelligence (AI) systems become increasingly integrated into critical domains, ensuring their responsible design and continuous development is imperative. Effective AI quality management (QM) requires tools and methodologies that address the complexities of the AI lifecycle. In this paper, we propose an approach for AI lifecycle planning that bridges the gap between generic guidelines and use case-specific requirements (MQG4AI). Our work aims to contribute to the development of practical tools for implementing Responsible AI (RAI) by aligning lifecycle planning with technical, ethical and regulatory demands. Central to our approach is the introduction of a flexible and customizable *Methodology based on Quality Gates*, whose building blocks incorporate RAI knowledge through information linking along the AI lifecycle in a continuous manner, addressing AI's evolutionary character. For our present contribution, we put a particular emphasis on the *Explanation* stage during model development, and illustrate how to align a guideline to evaluate the quality of explanations with MQG4AI, contributing to overall *Transparency*.

## 1. Introduction

The European regulation on AI, i.e. the EU AI Act, entered into force on August 1st 2024, and categorizes AI applications into four different levels of risk that lead to progressively stricter regulatory requirements, up to prohibition. The legal framework will come into effect in August 2026 [38, 1]. Overall, the Act directs towards trustworthy AI (TAI), which is ethical, lawful, and robust [6, 8] and ideally, results in intelligent systems that enter the European market respecting fundamental rights, safety, and health. [13] The *Assessment List for Trustworthy AI* [22], as published by an independent high-level expert group (HLEG) set up by the European Commission, provides a structured overview how to approach implementing TAI, organized into seven topics, including *Transparency* of intelligent systems, which we will highlight in more detail as part of the present contribution focusing on *Explainability* of deep neural networks (DNN). How DNNs generates output is not comprehensible for humans. The field explainable AI (XAI) comprises researching ways to shed light on the model's inner workings, and how to design methods towards human comprehensibility. Assessing the quality of generated explanations introduces novel challenges, as we explore in this paper. On an abstract level, intelligent systems share lifecycle processes, that describe all design decisions related to their transition from conceptualization to reality. However, multiple approaches exist, and a generalizable level is not trivial to organize for more specific design decisions and process stages, that depend strongly on factors such as data, type, and functionality of the intelligent system within a particular domain. As a result, AI challenges include its use case-specificity, stochasticity, and the black-box character of complex DNNs, which results in opacity by default. Currently, numerous resources are being published worldwide on how to implement Responsible AI (RAI) [9] systems. These resources include best practices, guidelines, and standards aimed at promoting auditability and accountability throughout the design, development, and use of AI systems, ensuring they meet the relevant specifications, ethical considerations, and regulations for their intended domain of use [9, 18]. Consequently, the question arises as how to organize RAI methods for implementing and regulating AI in the multitude of possible application scenarios to benefit society, which can only be measured in the real world. In the present paper, we outline our *Methodology based on Quality Gates* (MQG4AI) [10], [11], which follows principles from Design Science Research [5]. Our proposed approach towards realizing RAI focuses on AI system-specific lifecycle planning through information management (IM) [26] for individual AI projects, while simultaneously closing the gap to decentralized RAI design knowledge management (KM) [26] of on-going contributions, aiming to respond to the technique's evolutionary character. We illustrate the proposed method in form of a generalizable template for high-risk AI on GitHub[1] focusing on the model development stage and related risks. Based on customizable Quality Gates (QG) that mirror design decisions and process stages for information processing along

---

This work was partially funded by the German Federal Ministry of Education and Research (BMBF) under reference number 031L9196B, and partially supported by grant IAX from the Comunidad de Madrid Young Investigators initiative 2023/2024.

*Corresponding author

✉ miriam.elia@uni-a.de (M. Elia); albamaria.lopez@alumnos.upm.es (A.M. Lopez); k.cpaez@upm.es (K.A. Corredor Páez); bernhard.bauer@uni-a.de (B. Bauer); esteban.garcia@upm.es (E. Garcia-Cuesta)

ORCID(s): 0000-0001-6253-230X (M. Elia); 0000-0002-7931-1105 (B. Bauer); 0000-0002-1215-3333 (E. Garcia-Cuesta)

[1] https://github.com/miriamelia/MQG4AI

the AI lifecycle, we aim to relate implemented concepts with regulatory requirements, and ethics by design. Following the *NIST AI Risk Management Framework* [39], we focus on AI risk management (RM) structured according to AI trustworthiness criteria [22]. Further, we include related AI system-specific information, which is extracted based on the *AI Risk Ontology* (AIRO) that incorporates the AI Act and ISO 31000 series of RM standards [17].

The present contribution outlines our proposed template-design workflow to enable decentralized contributions, since finalizing the template is an immense endeavor, and we believe novel AI knowledge will never stop. First, we introduce approaches to AI QM in section 2, before outlining the methodological setup in section 3. Next, we transform existing best practices for the AI *Transparency*-related requirement *Explainability* [22] into AI risks, and derive corresponding QGs along the AI lifecycle in section 4. Finally, we illustrate our proposed QG information processing structure in section 5 for a technical guideline to implement a fidelity-robustness score that evaluates LIME-, and SHAP-explanations [30], addressing the risk *unfaithful explanations*. Section 6 discusses how MQG4AI is envisioned to support EU AI Act-conform QMS, before concluding our contribution in section 7.

## 2. Related Work

Intelligent systems must uphold and implement fundamental rights, safety, and health standards [13] when integrated into their intended real-world environments, which are inherently linked to their purpose and functionality. High-risk systems, in particular, pose significant threats to these principles if not properly designed and implemented. Managing AI risks presents challenges due to the use case-specific nature of these systems, their evolving characteristics, and the limited experience we currently have with their behavior in real-world contexts. Based on our use cases, we concentrate on the medical domain and related AI challenges, but we believe the approach is generalizable across domains, which is the focus of the present paper.

**Risk Management (RM)**: RM processes, as described in ISO 14971:2019 on the "Application of Risk Management to Medical Devices" [24], as well as risk concepts, as identified in the AI Risk Ontology (AIRO) [17],[2] can generally be summarized in *risk analysis*, *risk evaluation*, and *risk control* measures. They should align with the system's intended purpose and *reasonably foreseeable misuse*, which is defined as "[...] the use of an AI system in a way that is not in accordance with its intended purpose, but which may result from reasonably foreseeable human behaviour or interaction with other systems, including other AI systems" in Article 3 of the AI Act [13]. *Risk* is interpreted as "[...] the combination of the probability of an occurrence of harm and the severity of that harm" in Article 3 [13]. Accordingly, *risk sources* must be identified, multiple *events* derived that result in different levels of *harm*, and their *likelihood* estimated to ensure system *safety*, described in ISO 14971 as "freedom from unacceptable risk" [24, 6] during *risk analysis*. *Risk evaluation* organizes identified and estimated risk with respect to their impact within the concrete application context, and *risk control* measures mitigate risks. These steps are executed in an iterative manner, until *residual risk*, i.e. the risk that remains after implementing risk control measures, along with any new risks introduced by these measures, are sufficiently minimized according to "[...] criteria for risk acceptability defined in the [RM] plan" [24, 12]. Finally, a *benefit-risk analysis* [24, 14] can assist in evaluating *residual risk* when "[...] it is not judged acceptable [...] and further risk control is not practicable" [24, 14].[3] This involves comparing "[...] the benefits of the intended use [...]" [24, 14] against the *residual risk*. And ISO 14971 introduces *benefit* as "positive impact or desirable outcome of the use of a medical device on the health of an individual, or a positive impact on patient management or public health" [24, 2].

**AI Risks**: This abstract approach to RM extends to AI risks, focusing on "[...] those which may be reasonably mitigated or eliminated through the development or design of the high-risk AI system, or the provision of adequate technical information," as outlined in Article 9 of the AI Act [13]. AI RM needs to address the technology's inherent dynamics, i.e. its opacity, non-deterministic and evolutionary character, which results in a multitude of use case-adapted risk control measures along the AI lifecycle. Also, trade-offs between risk mitigating decisions can occur, e.g. between explainability of DNNs and their performance, which adds to the complexity of implementing RM for individual projects. The *AI Risk Repository*[4] offers a comprehensive foundation for risk identification, cataloging over 700 identified risks. Similarly, the *Mitre Atlas*[5] provides a "living knowledge base" focused on AI security and threats, enabling ongoing risk identification. Both repositories serve as valuable starting points for addressing known AI risks throughout the AI lifecycle. In addition, AI standards serve as knowledge base to identify and mitigate AI risks. The European Commission has released a document outlining the requirements for European standards. It highlights the development of standards in 10 directions: RM, data quality and governance, record keeping, transparency, human oversight, accuracy, robustness, cybersecurity, quality management, and conformity assessment [38].[6] Overall, "[...] standardisation should explicitly take into account the risks identified as part of the [RM] process" [38, 4].

**Interoperability in AI RM**: The Organisation for Economic Co-operation and Development (OECD) published a document where they compare different approaches to AI

---

[2]https://delaramglp.github.io/airo/, accessed December 2024.

[3]The concept of *acceptable risk* needs to be addressed carefully. The *Medical Device Regulation (MDR)*, which is harmonized with the AI Act, as stated in Annex I [13], emphasizes "[...] the reduction of risks as far as possible without adversely affecting the benefit-risk ratio" [32] in Annex I, Chapter 1.

[4]https://airisk.mit.edu/

[5]https://atlas.mitre.org/

[6]For an overview of European standards and their status, visit: https://www.cencenelec.eu/european-standardization/european-standards/

RM, among others, the EU AI Act's Article 9 (as published in 2023), ISO/IEC 23894 *AI - Guidance on RM* (which extends ISO 31000 *Risk Management - Guidelines* with AI-specifics) and the NIST *AI Risk Management Framework* (RMF) [34]. They found these approaches "[...] are generally aligned with four top-level steps: 'DEFINE', 'ASSESS', and 'TREAT' risks, and 'GOVERN' [RM] processes" [34, 10]. And, aiming for accountability necessitates "[...] to follow these steps at each phase of the AI system lifecycle [...]" [34, 16]. The first three steps comprise the previously introduced RM processes in alignment with application-specific contextual information [34, 10]. For instance, the NIST AI RMF emphasizes adapting risks to real-world settings, transparent documentation [39, 6], involving a "[...] broad set of perspectives and actors across the AI lifecycle" [39, 9] to ensure comprehensive risk analysis, and continuous RM throughout the AI lifecycle [39, 20]. Most "[...] differences between frameworks relate to the 'GOVERN' function" [34, 11], which describes relevant steps how to embed RM processes within the organization. This includes monitoring, and documenting the RM process, as well as the assignment of responsibilities, among others. [34, 10] The NIST AI RMF, as well as ISO standards map closely to the OECD's interoperability framework, while the AI Act stands out through the four-level risk classification and the associated emphasis on specific requirements for high-risk applications. Further, it seems to be missing certain GOVERN RM measures (i.e. stakeholder consultation and integrating RM into the organizational culture). [34, 11] In this regard, the AI quality management system (QMS) "[...] comes closest to embedding the [RM] system into broader organisational governance" [34, 27].

**AI Quality Management (QM)**: RM contributes to AI QMS, as specified in Article 17 of the AI Act [13], and "[t]he required AI QMS is the foundational building block to ensure ongoing quality and compliance [...]" [1, 2]. We discuss the MQG4AI concept and the mandatory AI QMS in section 6. It comprises 13 aspects, among which the AI RMS, as described in Article 9. In addition, a strategy for regulatory compliance; design, design control and verification; development, quality control and assurance; continuous examination, test, and validation procedures; technical specifications; data management; a post-market monitoring system (Article 72); reporting of serious incidents (Article 73); communication with national competent authorities; record keeping; resource management; and an accountability framework are required [13]. An overall orientation how to design and integrate AI QM within the organization is outlined in ISO/IEC 42001 on *AI Management Systems* [27], which is "[...] not aligned in objectives and approach with the AI Act [...]" [38, 6], but provides "[...] some relevant clauses at the technical and organizational levels. These could be referenced, as appropriate, by new standardization in [QM] for AI, while ensuring that its focus remains on the specific risks and objectives captured in the legal text" [38, 6]. In addition to information on policy, resource and documentation management, the standard equally highlights the role of RM and the AI lifecycle [27]. While offering comprehensive conceptual guidance for addressing RAI, the concrete implementation of recommendations, such as to consider "life cycle stages [...]; testing requirements and planned means for testing; human oversight requirements [...]" in processes for the responsible design and development of AI systems [27, 31], or "testing methodologies and tools; selection of test data and their representation of the intended domain of use; release criteria requirements" for AI system verification and validation [27, 32], remains relatively high-level. Finally, beyond standards, more or less comprehensive frameworks and approaches that contribute knowledge to parts of AI QM exist, reaching from documentation management, or (domain-specific) evaluation approaches, over software requirements in the AI context to the implementation of ethical principles [18] [15] [36] [37] [20].

**Need for AI QM Tools**: In summary, the concrete implementation of AI QMS criteria [13], which must account for the technology's evolutionary and non-deterministic nature, is a non-trivial task. Probably thanks to this complexity, compared with publications on AI risks, and the reliable evaluation of AI components, "[...] there is less research on designing and implementing a quality management system (QMS), [...] and no clear suggestion exists [...]" [33, 2]. As a result, implementing comprehensive "practical tools" [33, 1] that support AI QMS processes and requirements that go beyond scientific prototypes [37, 11] is needed to support the implementation of the AI Act, while fostering AI innovation, contributing to the work of the AI Office [1, 4], which is tasked with implementing the AI Act. Overall, when designing such tools, "the integration of them across the life cycle to provide a holistic system is critical for wide scale use" [37, 11]. In [33], for instance the authors propose a microservices-based AI QMS tool that is envisioned to enable user-guided quality checks of AI system components along the AI lifecycle, as well as automated generation of the required documentation. They provide a first prototype focusing on RM for large language models (LLM). Similar to MQG4AI, as outlined in the following sections, they aim for flexible components and comprehensive lifecycle integration. Possibly, elements of both proposed methods could be combined for tool development. To the best of our knowledge, no comparable approach to MQG4AI exists to date, while conceptually similar procedures, that build on QGs and information management (IM) can be found in the context of traditional product development [16] [19]. Addressing AI's inherent dynamics, MQG4AI focuses on continuous AI lifecycle planning through IM, as well as continuous RAI knowledge management (KM). Overall, we aim to contribute a generic and customizable methodology, that could be implemented as a tool, towards facilitating AI QM, as explained in the following sections.

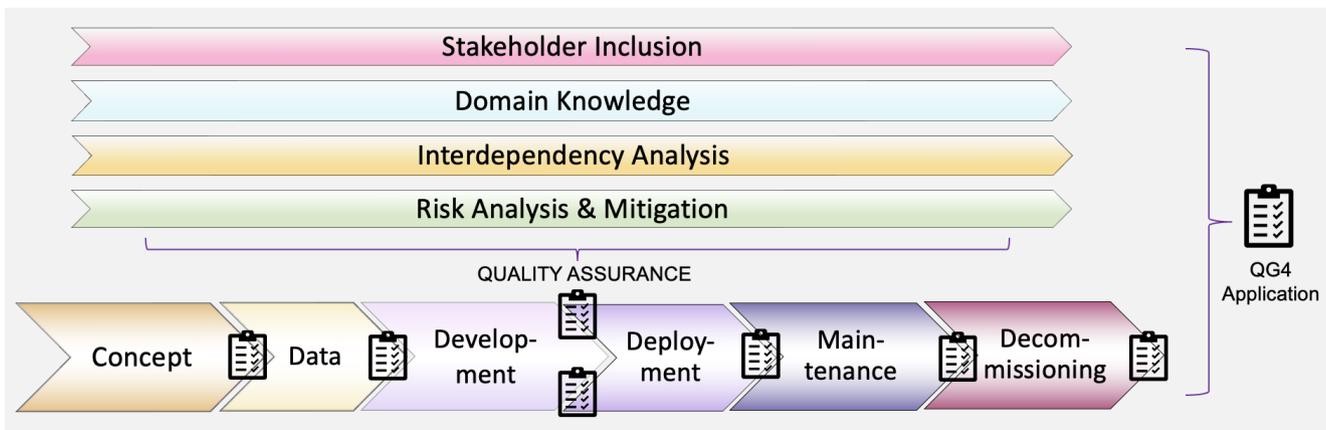

**Figure 1:** The most generalizable Quality Gates (QG) along the AI lifecycle are depicted, their creation and realization follow four fundamental principles towards RAI. The extracted process steps are to be understood in an iterative and intertwined manner for implementation, which is envisioned to be integrated with MQG4AI by design through information management and evolving template versions of the AI lifecycle. Our depiction follows an application-oriented view, aiming to capture concepts for implementation: Where is the model and what does it need?

## 3. MQG4AI Setup

This section introduces our proposed Methodology based on Quality Gates (MQG4AI) towards implementing Responsible AI (RAI) [9]. In addition to traditional software engineering practices, AI introduces new elements, which should be maintained in a *backlog*, supporting "[...] cross-disciplinary coordination, planning and evaluation" [26, 26], which is the focus of the MQG4AI-template. Concretely, we envision to provide an information management (IM) structure that contributes to AI Act-conform Quality Management Systems (AI QMS) (Article 17) from the development perspective through comprehensive lifecycle planning at the interface with compliance assessment. It is the provider's responsibility to prove the required quality, and, among other requirements, AI QMS include a risk management system (RMS) (Article 9) which is the starting point of MQG4AI design. In addition, the Act highlights the crucial role of AI literacy in Article 4, which necessitates comprehensive RAI knowledge management (KM). [13] ISO 5338:2023 on *AI system lifecycle processes* defines the lifecycle IM process as follows:

> The purpose of the [IM] process is to generate, obtain, confirm, transform, retain, retrieve, disseminate, and dispose of information, to designated stakeholders. The [IM] process plans, executes, and controls the provision of unambiguous, complete, verifiable, consistent, modifiable, traceable, and presentable information to designated stakeholders. Information includes technical, project, organizational, agreement, and user information. Information is often derived from data records of the organization, system, process, or project. [26, 16]

Aiming to close the gap to compliance for high-risk AI, we propose structured IM, aligned with evolving RAI knowledge along the AI lifecycle for contributing stakeholders that play an active role when implementing lifecycle processes. KM processes in the context of MQG4AI are depicted in Figure 2, and ISO 5338 defines KM as follows:

> The purpose of the [KM] process is to create the capabilities and assets that enable the organization to exploit opportunities to reapply existing knowledge. This encompasses knowledge, skills, and knowledge assets, including system elements. [26, 11]

In summary, the MQG4AI-template is envisioned as a tool for contributing stakeholders to interpret project-specific lifecycle information through a generic and customizable format that incorporates continuous RAI KM by design.

MQG4AI's four fundamental design principles are illustrated in Figure 1.[7] The following sections introduce our proposed MQG4AI-template, as illustrated on GitHub. First, the template sections are introduced, envisioned to summarize supplementary contextual information that is linked with the AI lifecycle conceptualization. Next, the two MQG4AI interaction scenarios, based on Design Science Research (DSR) [5] are outlined, aiming to respond to the evolutionary character of AI knowledge, before detailing QGs along the AI lifecycle, which comprise the core of the MQG4AI-template.

### 3.1. Template Sections

The proposed MQG4AI-template is envisioned to provide a unifying approach for comprehensive lifecycle planning towards implementing RAI. We operate under the assumption that every lifecycle design decision contributes to ensuring the desired quality of AI systems, which is summarized in form of QGs. They comprise the center of MQG4AI, and design the lifecycle from generalizable to use case-specific implementation guidelines. Their generic and customizable structure, designed to address the dynamic nature of AI and to integrate design decisions with relevant contextual information, is elaborated in more detail in the following sections. Focusing on RM as core-requirement towards AI QM, the current version of MQG4AI consists of the three template-sections *System*, *Risk Management*, and *Lifecycle* information. Overall, the proposed MQG4AI information and building blocks are designed to offer flexibility in their definition and arrangement, aiming for interchangeability and docking with existing information gathering, such as medical QM based on ISO 13485 [23], which is expected to function as foundation for medical AI QM [1, 2], for instance. The extracted information flow between the proposed sections is illustrated in Figure 3.

---

[7]Figure 1 is a refined version of the lifecycle graphic in [10, 4], where we introduce basic methodological building blocks.

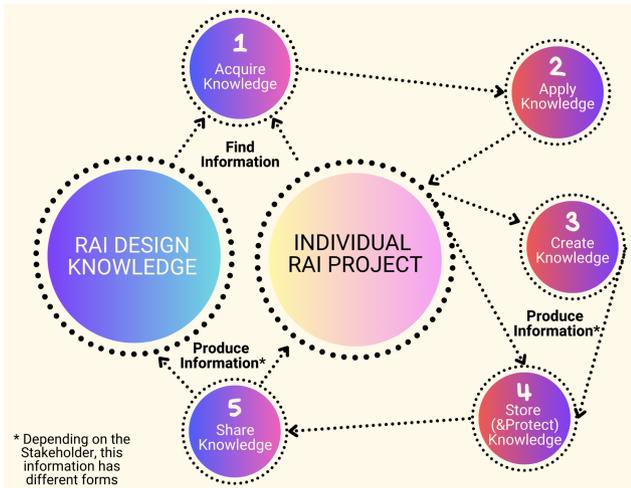

**Figure 2:** Knowledge Management (KM) processes [3, 177] in the context of MQG4AI are integral to the developer's perspective on Responsible AI (RAI) knowledge and lifecycle design decision-making. Developers acquire project-specific lifecycle knowledge through the analysis of existing information [26, 21], which is subsequently applied and implemented as lifecycle design decisions. Throughout the lifecycle execution, new knowledge is generated within the context of the project, documented, and stored using the MQG4AI-template for lifecycle planning, encompassing the applied knowledge. The template also facilitates knowledge sharing within the AI project and externally, contributing to broader AI literacy. This iterative exchange between information and knowledge aims to iteratively refine lifecycle design, ensuring the desired quality of the AI system is achieved.

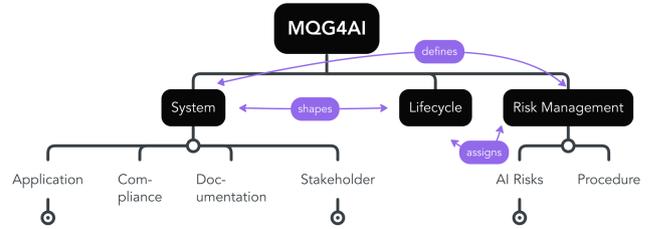

**Figure 3:** Proposed bidirectional information flow to link the AI lifecycle implementation with supplementary contextual information.

**AI System Information** The integrated AI system-information structure is based on the AI Act-conform AIRO [17], and extracted information is envisioned to be exchangeable with existing QM procedures that possibly already collect the required information. For our contribution, we focus on the extraction of interdependencies with lifecycle design decision-making, such as application-specific content that affects the implementation (as well as RM), which in turn comprises relevant information for the post-market monitoring system that needs to be documented. We identified five sub-sections for system information based on the AIRO webtool as a starting point: *Application*, *Compliance*, *Documentation*, *Ethics_General* and *Stakeholder*, which are explained on more detail in the MQG4AI template.[8] In addition to AIRO, we append *Ethics* sections, complementing general information on ethics in the context of RAI, *Ethics_Specific* is tailored to the respective application, and the *Domain Knowledge* subsection aims to provide relevant knowledge on the concrete *Application* to all stakeholders, following MQG4AI's four fundamental design principles, as depicted in Figure 1. Further, we propose to structure stakeholder roles into *active*, *consulting*, and *passive* for directed IM, since the stakeholder's involvement with the project impacts how information is tailored and presented to them.

**Risk Management Information** Further responding to regulatory requirements, we focus on the identification of generalizable AI risks that are related with implemented lifecycle design decisions structured according to AI trustworthiness (TAI) criteria. The NIST AI RMF highlights the relationship between negative risk reduction and AI trustworthiness [39, 12], which is equally included with the OECD's RM interoperability framework [34, 19]. Other approaches to RM are possible, and we are convinced that all AI risks, which are based on AI pitfalls [21] [31], for instance, are alignable with the generic ALTAI structure.

The seven TAI criteria comprise: *Human agency and oversight*, *Technical robustness and safety*, *Privacy and data governance*, *Transparency*, *Diversity, non-discrimination and fairness*, *Societal and environmental well-being*, *Accountability* that are intended to be addressed along the complete AI lifecycle through technical and non-technical methods [22]. Note their similarity with directions for standard development, as discussed for AI risks in 2, while the key areas place greater emphasis on methods for implementing TAI criteria. This focus is particularly evident in conformity assessment and QM, where the TAI criteria of fairness and societal well-being are addressed indirectly through the evaluation of requirements the intelligent system must meet to achieve its intended purpose. [38]

---

[8] https://github.com/miriamelia/MQG4AI

Finally, risks are related with system-information allocating the required documentation, among other information flow scenarios, such as application-specific information that equally shapes RM. We aim to advance from abstract concepts to use case-specific RAI design decisions, enabling comprehensive AI lifecycle planning using the MQG4AI-template. This is achieved through qualitative, RAI knowledge-based IM to support lifecycle planning and monitoring, ensuring accessibility for contributing stakeholders. Therefore, a structured and generic method for AI risk identification is included within the MQG4AI-template, ensuring continuous and thorough coverage throughout the AI lifecycle implementation. Overall, we simulate linking design decisions with AI risks, and supplementary information blocks are exchangeable with existing, or differently gathered information.

## 3.2. Interaction Scenarios

Aiming to address AI's evolutionary character and to promote AI literacy, MQG4AI-template design is based on principles from Design Science Research (DSR) [5], i.e. the continuous communication between an abstract design knowledge base, and concrete use cases to test/update/transform the accumulated, and continuously evolving knowledge on RAI design. This results in two differing MQG4AI-template interaction scenarios: *public Design Knowledge* (MQG4DK), and *private Application* (MQG4A).

**MQG4DK** The MQG4DK-template is intended to enable summarizing the ever evolving and dynamic RAI design knowledge along the AI lifecycle. This is based on generic and customizable QGs through information processing of practical guidelines that include clear instructions and are related with supplementary AI QMS information. We begin with AI system-specific and AI risk information and aim for customizability of additional information transformation and extraction. MQG4DK is envisioned to grow in a decentralized manner, and we intend to design the foundation for further contributions based on the extracted template design workflow, while providing customizability for connecting other reasonable QM information blocks. Overall, MQG4DK contributes towards continuous RAI lifecycle design knowledge organization.

**MQG4A** MQG4A is envisioned to be applied as an additional management layer for AI knowledge-based lifecycle conceptualization and management during individual AI projects, as promoted by the European AI Office for RMS, including criteria for acceptable methods.[9] We concentrate on providing an information management (IM) tool that is accessible to contributing stakeholders along the AI lifecycle. The envisioned workflow is organized similarly to the Git branching structure that enables decentralized contributions to a central project, which undergo quality checks before a final merge. During project application, different MQG4A-versions include concrete results, and they are intended to reflect the evolution of individual lifecycles. For instance, MQG4A-versions can be structured into pre-, intra-, and post-selection versions for lifecycle planning. This idea is based on our empirically extracted design decision-making method towards reliable performance evaluation metrics for a fictional use case situated in emergency medicine [11], and the proposed structure is extendable to establish relevant design decisions during model development. Emphasizing interdependencies that impact model optimization, the iterative process of design decision-making is illustrated in Figure 4, where we highlight generated results and their updates for the interrelated model configuration, explanation, and preprocessing steps that are built on data quality, as well as a solid performance evaluation strategy.

**MQG4A & MQG4DK Interaction** Communication of RAI lifecycle knowledge between these two template perspectives happens through QGs, as outlined in the next section in more detail. The use case-adaptation of MQG4A is envisioned to be realized by a configurable pull version from MQG4DK (MQG4A-v0). The QG-format provides information layers that enable an intelligent search through use case-specific attributes (see leaf-QG-naming and -tags). As a result, only applicable lifecycle design decisions are pulled from the design knowledge base, in addition to the generic default template for high-risk systems. With respect to MQG4DK, MQG4A applies/provides/updates/transforms RAI knowledge adapted to the use case at hand in an optimized form, compared with other MQG4DK-contributions through a shared format by default.

## 3.3. AI Lifecycle Design via Quality Gates (QG)

Quality Gates (QG) along the AI lifecycle are at the core of the MQG4AI information structure, drawing from traditional software engineering and product development practices. Generally, QGs "[...] structure a process chain into phases and allow a periodical review of the process quality" [19, 206], reflecting a decision point within a project [14, 245], and ensuring consistent assessment and control of quality at various stages. We adapt the basic concept, which entails QG-identification and positioning, as well as the definition of desirable QG-criteria to the AI context. [12] [14] [16] [19]

Therefore, we design a generic and customizable information structure based on QGs that incorporate AI-specific dynamics, aiming for quality by design. Overall, we differentiate two general types of QGs for information structuring and processing. Leaf-QGs mirror design decisions, and collection-QGs construct the AI lifecycle. Combined, they can be interpreted to extract interdependencies in form of a project-specific QG lifecycle graph structure with (bi-)directional vertical and horizontal information connections. Following the lifecycle flow [26, 36], as defined by ISO/IEC FDIS 5338 on *AI system life cycle processes* [26], and inspired by the authors of [8], who emphasize AI application-specific lifecycle design, we aim to provide a basic high-level lifecycle structure that is generalizable for the multitude of AI lifecycle design choices. Overall, we design the lifecycle in a way that ensures applicability across a wide range of use case-specific implementations, focusing on documenting and monitoring implemented concepts for continuous planning. Concurrently, we aim to provide a structure that accommodates differing providers for specific lifecycle stages. For instance, the *data* can be bought externally for in-house model *development*, as well as particular *deployment*, and *maintenance* services, while sub-processes, such as *data preprocessing*, or the intelligent system's *onboarding strategy* in its intended real-world setting need to be developed in-house, in alignment with the intended purpose.

---

[9]Refer to this video from May 2024, outlining the intended RMS logic for more information: https://webcast.ec.europa.eu/risk-management-logic-of-the-ai-act-and-related-standards-2024-05-30?utm_source=substack&utm_medium=email

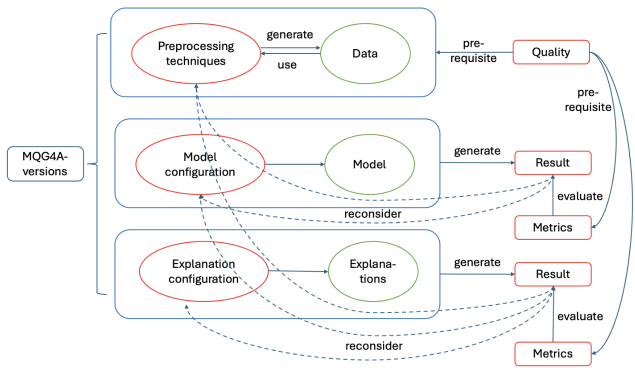

**Figure 4:** The iterative process of configuring the model (including hyperparameters) and explanation method is illustrated for the optimization step during the development stage, highlighting outputs and interdependencies. In addition, these design decisions are influenced by the preprocessed data, and evaluated against predefined metrics that need to be suitable for the respective context and structural setting. Their definition belongs to different stages and needs to be planned accordingly. A combination of MQG4A-versions is envisioned to structure the iterative process of design decision-making as part of a continuous lifecycle conceptualization. For instance, MQG4A-versions can be structured into pre-, intra-, and post-selection steps. Based on the required information identified prior to considering the design decision, such as evaluation metrics, evolving results are compared with modifications to a baseline approach until reasonable updates are identified.

**Generic High-level AI Lifecycle Stages** As illustrated in Figure 1, we identify six generalizable phases (*Conceptualization*, *Data*, *Development*, *Deployment*, *Maintenance*, *Decommissioning*), whose execution should be understood as iterative. Each lifecycle evolution is characterized by a distinct focus, depending on the project's progress, while all stages need to be considered from the start of project *Conceptualization*. The MQG4AI-template is envisioned as a lifecycle planning tool for iterative conceptualization stages, and different MQG4A-template versions comprise information on the lifecycle evolution. In addition, the option to derive MQG4A-template branches can be applied to monitor information on combinations of design choices for conducted analysis regarding system validation, for instance. Our generic high-level structure globally aligns with the OECD's lifecycle [34, 16], which mirrors ISO/IEC 22989 *Artificial intelligence concepts and terminology* [25, 36] (on which ISO/IEC 42001 on *Artificial intelligence management system* is based [27, 31]), while a few differences emerge based on MQG4AI's lifecycle planning functionality. Concretely, we aim to address the questions *Where is the model and what does it need* for the proposed generic lifecycle definition, with the goal of capturing concepts for practical implementation. For instance, the proposed structure equally aligns with the Machine Learning Operations (MLOps) lifecycle-view, as outlined in [35, 3].

1. We include *Inception*, i.e. the definition of contextual information such as business analysis or system requirements [26, 6] within the *Conceptualization* phase. This information, once the project's feasibility is determined, fills MQG4AI's customizable contextual information blocks for information linking.

2. Further, we separate the *Design* phase from *Development* to highlight continuous lifecycle *Conceptualization*. During *Conceptualization*, planning and continuous updates centered around the model and in alignment with findings from other stages and contextual information is conducted based on the MQG4AI-template.

3. Next, we explicitly highlight the *Data* stage which comprises design choices and information for further lifecycle design decisions. There, the model may not yet exist or is in a state of update, with the quality and transformation of data continuously influencing the model's quality.

4. During *Development*, the model is created, evaluated and refined. In contrast to ISO 22989, *Verification and Validation*, i.e. testing if the developed model works as expected [25, 39] is executed through different MQG4A-template versions that mirror conceptual combinations including concrete results in alignment with the previously identified metrics based on *Development* and related information. For instance, qualities such as model robustness are tested with adversarial examples that comprise changes in the data stage during preprocessing and impact performance evaluation. Therefore, this information is summarized as a MQG4A-template branch and not merged into the "final" model concept that is released into the real world.

5. *Deployment* and *Maintenance* focus on integrating and monitoring the model in real-world scenarios. These stages are interconnected with *Development* and *Data*, while introducing additional information layers dictated by the target environment. This includes real-world data distributions and integration with existing pipelines, highlighting MLOps. *Continuous Validation* and *Re-evaluation* of model outputs in the real world, as highlighted by ISO 22989 [25, 40] and refined by ISO 5338 [26, 5], who emphasize that *Continuous Validation* "[...] is also applicable in situations without continuous learning, for example, to detect data drift, concept drift or to detect any technical malfunctions" [26, 5], contribute to designing the *Maintenance* stage. They are influenced by identified conceptual interdependencies with the *Development* and *Data* stages. The evaluation of generated results may possibly restart a new lifecycle planning cycle, building on the previously identified knowledge that can be transferred to the next MQG4A-version.

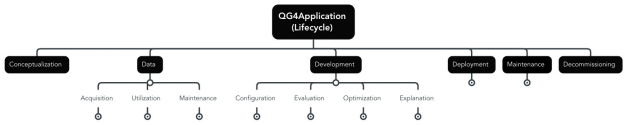

Figure 5: Proposed sub-level of generalizable collection-QGs for the *Development* and *Data* stage.

6. Finally, *Decommissioning* addresses relevant considerations for the model's retrieval from real-world applications, potentially reversing steps from previous stages. ISO 22989 includes model replacement with an advanced version with the *Retirement* stage [25, 40], which we include within the *Maintenance* stage.

This approach facilitates the integration of conceptual information on AI projects within a comprehensive AI system lifecycle, systematically organized across multiple sub-levels. Based on our experience, we concentrate on abstracting generalizable process steps for the *Development* stage as starting point for MQG4AI design. Focusing on the *Explanation* stage, related *Collection-QGs*, as depicted in Figures 5, and 6 are outlined in more detail as part of our present contribution in section 4.

**Collection-QGs** Collection-QGs are hierarchically structured, and start from the most generalizable level of the AI lifecycle to summarize and organize their related sub-QGs, aiming to close the gap to use case-specificity. They can include up to *n* sub-QG levels that are either collection- or leaf-QGs. The former construct the QG-graph and identify vertical interdependencies that are envisioned to be generalizable across use cases for higher levels along the AI lifecycle, as illustrated in Figure 5. Their respective leaf-QGs provide more use case-specific information tailored to specific AI techniques, as depicted in Figure 6. Finally, a single root node, i.e. *QG4Application*, summarizes information on all lifecycle-QGs. Overall, interpreting AI lifecycle information from different vertical levels can be applied for customizable system evaluation scenarios in form of a QG-scoring system. For instance, a *stakeholder-participation score* could be derived based on a vertical and horizontal analysis of stakeholder inclusion across all QGs within a pre-defined collection-QG scope. Stakeholder participation contributes to the fifth ALTAI criterion *Diversity, Non-Discrimination and Fairness* [22, 18], and a lack of domain experts, for instance poses risks.

**Leaf-QGs** Focusing on design decision-making along the AI lifecycle, leaf-QGs compose an information template that is intended to provide the necessary channels that ensure the desired quality through a customizable information format and linking of interdependent information, as depicted in Figure 7, and exemplified in section 5. Our starting point for defining the leaf-QG template responds to the on-going conceptualization of the lifecycle *Development* stage, broaching the strongly related *Data* stage, and including information extraction for *Deployment*, as well as *Maintenance*. The template is extracted based on a retrospective analysis of interdependencies during model development focusing on reliable performance evaluation [11]. If not implemented properly, unreliable performance metrics pose risks [21] that are propagated along the AI lifecycle, since they function as the foundation for further design decision-making. At their core, leaf-QG creation is based on a three-fold structure (*Content* definition, *Method* extraction, and content *Representation* tailored to stakeholder-views),[10] which is related to relevant input, and output information. We aim to enable a comprehensive interdependency analysis, and the proposed information structure simultaneously contributes to testing by design of acceptable implementation approaches through a multi-perspective transformation. Further, additional information layers are intended to close the gap to overall IM towards a RAI application, which we illustrate for TAI RM. The format can be adapted to other stages and types of information, with the template intended to offer sufficient customizability, making it scalable to include information on AI QMS requirements.

1. *Input Information* comprises related AI system information regarding the intended use or required stakeholder roles, for instance, as well as other lifecycle QGs that provide necessary information for individual leaf-QG creation.
2. *Output Information* extracts AI system-information, e.g. for post-market monitoring, and links other related lifecycle QGs that are impacted.
3. The three *Leaf-QG Creation* dimensions are envisioned to ensure the desired quality of lifecycle design decisions through information processing, which we illustrate in section 5 in more detail.
4. The *Evaluation* layer outlines open questions and possible draw-backs of the selected approach to implement a design decision, which can function as foundation for (residual) risk analysis.
5. The *Risk Management*-layer relates individual design decisions with identified risks to structure the continuous implementation of risk controls and repeated (residual) risk analysis as part of MQG4A-versions.
6. The *QG-Naming*-structure is intended to link information with particular AI techniques such as classification or segmentation scenarios, as well as to identify horizontal lifecycle interdependencies of design-decision contributions that are applicable to a shared structural setting.

---

[10]We focus on information extraction for the *Content* and *Method* dimension during development, and illustrate the *Representation* dimension.

It is broadly defined as **QG_name_(view)**, while *name* describes the leaf-QG content, and *view* details the AI technique where the respective QG is applicable to fill MQG4DK. A practical realization of *view* for the MQG4A-template is part of future work, possibly it could contribute information on the template version, which mirror lifecycle-iteration cycles. For instance, the lifecycle collection-QG that comprises approaches to evaluate the quality of explanations has different implementations depending on the applied XAI method, which results in multiple possible leaf-QG contributions to MQG4DK. Consequently, the in section 5 introduced fidelity-robustness score is named: *QG_FidelityRobustnessScore_(SHAPLIME)*

7. *QG-Tags* are relevant to identify use case-specific lifecycle-QGs to pull a suitable MQG4A-template through a configurable version and an intelligent search. The supplementary contextual information sections, as well as high-level generalizable lifecycle collection-QGs are part of the default high-risk MQG4AI-template, and included in MQG4A-v0. The proposed information structure to tag leaf-QGs, is based on design patterns for machine learning applications, as proposed in [40]: **Name, Intent, Problem, Solution, Applicability, Consequences, Usage Example**

In summary, the described generic and customizable IM format is envisioned to enable a decentralized merging of generated knowledge on RAI lifecycle implementations for the multitude of possible AI use cases. This is realized through the continuous interplay between abstract MQG4DK and applied MQG4A. The premise of our approach is that AI use cases are grouped based on structural similarities and domain-specific conditions. Long-term, the MQG4AI template is envisioned to enhance the selection of existing methods that address AI risks, and aim to avoid pitfalls, while documenting the (individual) lifecycle concept for contributing stakeholders to enable a comprehensive and continuous lifecycle planning phase. MQG4DK-contributions are transferable to a specific AI use case based on an intelligent tag-search. QGs for MQG4A are created and updated during the conceptualization stage to organize, implement, and monitor AI lifecycle-relevant design decisions in a responsible and continuous manner. After introducing the MQG4AI concept, in the next section, we illustrate a possible workflow to determine lifecycle-QGs based on identified best practices related to *Explainability* for MQG4AI-template design.

## 4. QGs to Design the AI Lifecycle: From AI Risks to Best Practices and Risk Controls for Model Explanations

Deep neural networks (DNN) inherently lack transparency. As a result, explanation methods must be implemented in a responsible manner along the AI lifecycle. The field explainable AI (XAI), for instance contributes approaches how to implement this requirement. The present section first introduces two identified risks surrounding model opacity that mirror *Usability* and *Quality*, which comprise the generic directions for explanation evaluation. Related to these risks, we illustrate how to implement a set of three best practices along the AI lifecycle using MQG4AI's building blocks towards risk mitigation by design. Finally, we outline how our proposed application-oriented *Explanation* lifecycle structure corresponds to the *IEEE Guide for an Architectural Framework for Explainable Artificial Intelligence* [7]. The next section details the leaf-QG template and aligns a concrete design decision for the evaluation of explanation quality within the proposed application-oriented lifecycle structure. The resulting risk mitigated AI lifecycle explanation stage within the MQG4AI template can be viewed on GitHub.[11]

### 4.1. Explainability Risks & Best Practices

We identified two major explainability-related risks that we align with the MQG4AI information structure centered around the lack of model transparency for DNNs: (1) incomprehensible explanations to the user, (2) unfaithful or unreliable explanations regarding their quality, which is linked with two of the identified best practices.

1. *Incomprehensible Explanations to the User*: This risk arises when the methods of explanation and their presentation are not adapted to the user's capacity of comprehension confusion and misunderstanding. Explanations that are too vague, overly complex or are not well framed may prevent the user from understanding the model's reasoning on how to apply the results in a particular context, which may reduce trust in the system.

2. *Unfaithful or Unreliable Explanations*: This risk arises when explanation methods fail to accurately reflect a model's true reasoning, leading to misunderstandings of its decisions. Such explanations can misguide stakeholders, resulting in harmful actions throughout its lifecycle and eroding trust in the system.

The flip side of these risks are best practices, which support designing the intelligent system in a way that automatically mitigates risks. Linking best practices and risks highlights the interdependencies for monitoring risk controls. We aim to demonstrate how RM can be integrated with the AI lifecycle implementation, enabling all design decisions to be jointly evaluated and adjusted in alignment with the dynamics of AI lifecycle evolutions. Information linking is further conducted through the leaf-QG template information structure for concrete design decisions, which is explained in the next section.

1. *Comprehensible Explanations of Model Output to the User*: To foster a trustworthy application, explanations must be clearly communicated to the respective user. This includes the development of informative interfaces, at different levels of complexity, further detailing the explanations to facilitate user interpretation. In addition, a user interaction flow needs to be designed to ensure the intended understanding of the system's functioning, including the assertion of knowledge about the system's contextual operation and limitations. It should include a notification protocol when there are relevant updates that may imply changes in the AI system's operation. Concrete implementations have multiple forms depending on the use case. For instance, a pop-up notification window could appear informing the physician of potential inaccuracies, when an intelligent clinical decision support system (CDSS) encounters data from a female patient, but was mainly trained on data from male patients.

---
[11] https://github.com/miriamelia/MQG4AI

2. *Preferred Use of Interpretable Models (Ante-hoc)*: The optimal way to address the risk of unreliable explanations is to use ante-hoc explainability methods, if possible for the use case at hand. Ante-hoc models are intrinsically interpretable models designed to provide clear insights into their decision-making processes without relying on post-hoc explanation techniques. These models include models with simple decision boundaries, such as decision trees, and more complex ones with interpretability enhancements, like neural networks with prototype layers. For example, a neural network with a prototype layer may classify inputs by comparing them to representative prototypes, ensuring that predictions are grounded in tangible, human-understandable references [29]. By integrating interpretability into the model's architecture, ante-hoc models reduce the risk of unreliable explanations, as inherent explanations are more representative of the model's decision-making process. Therefore, required evaluation is mostly in terms of model performance.

3. *Validate the Explanations*: Finally, to address the risk of unreliable explanations, it is essential to validate explanations in terms of both their desired mathematical properties and their effect on the user's understanding. This process ensures that explanations are not coincidental or misleading, but instead are meaningful, consistent, and trustworthy. Moreover, the desirable properties of an explanation vary depending on the specific explanation technique employed. For example, post-hoc explanations generally require more nuanced evaluation than ante-hoc models; and example-based explanations should be evaluated for representativeness, relevance, and similarity, whereas for rule-based explanations properties as completeness, consistency, and conciseness are key.

After defining relevant criteria for the integration of explanations with the AI lifecycle, we construct the *Explanation* lifecycle stage focusing on the organization of reliable implementation concepts. As a result, on an abstract level, our proposed lifecycle definition workflow combines risks and best practices with the AI lifecycle towards quality by design.

### 4.2. Towards a Reliable Explanation Stage during Development: Proposal of Risk Controls by Design

Among the proposed generic MQG4AI lifecycle structure, as previously introduced in section 3.3, we study the generation of the *Explanation* lifecycle stage in more detail within the present paper. Aiming to close the gap between use case-specificity and broad applicability, we propose a combination of collection- and leaf-QGs that are intended to be generalizable across use cases for MQG4A-scenarios. Simultaneously, we offer a structure to assign use case-specific guidelines on lower hierarchical levels to MQG4DK.[12] Figure 6 portrays the proposed *Explanation* lifecycle stage, with three main stages: *Configuration*, *Evaluation*, and *User Interaction*.

**Leaf-QG** *Configuration* refers to a summary of relevant metadata and explanation characteristics that apply to any explanation method. With respect to application, they define the setting for one XAI method, and multiple methods can be appended. As seen in Figure 6, XAI method definition includes factors such as purpose, applicability, scope, result, and stage, with different options specified for each. Ideally, these options cover all potential explanation methods, providing a comprehensive framework to guide the selection of the most suitable approach.

- *Purpose* refers to the reason why explanations are generated, which is critical for risk mitigation. Developers can annotate whether explanations are needed to validate the model, assess data preprocessing techniques, inform stakeholders about model decisions, or uncover new insights learned by the model. This information helps in selecting the appropriate evaluation strategy for the explanations.

- *Applicability* indicates whether the explanation method is model-agnostic (applicable to any machine learning model) or model-specific (designed for a particular type of model). This decision affects both computational resources and model flexibility. Model-specific methods may require more computational power during implementation but could be more efficient once deployed, while model-agnostic methods are more flexible but potentially more computationally expensive.

---

[12]Note that the structure broadly resembles the proposed sub-level for the *Development* lifecycle stage, as depicted in Figure 5. However, the explanation method's configuration is defined as a leaf-QG, while model configuration comprises more complex sub-steps and is therefore, defined as a collection-QG. Also, the explanation *Optimization* stage is conducted through different MQG4A-template versions that address different implementations of input information interdependencies, such as data pre-processing and the underlying model, as depicted in Figure 4, since there are no specific explanation optimization methods. In contrast, model *Optimization* is additionally characterized by individual sub-steps, e.g. post-processing methods.

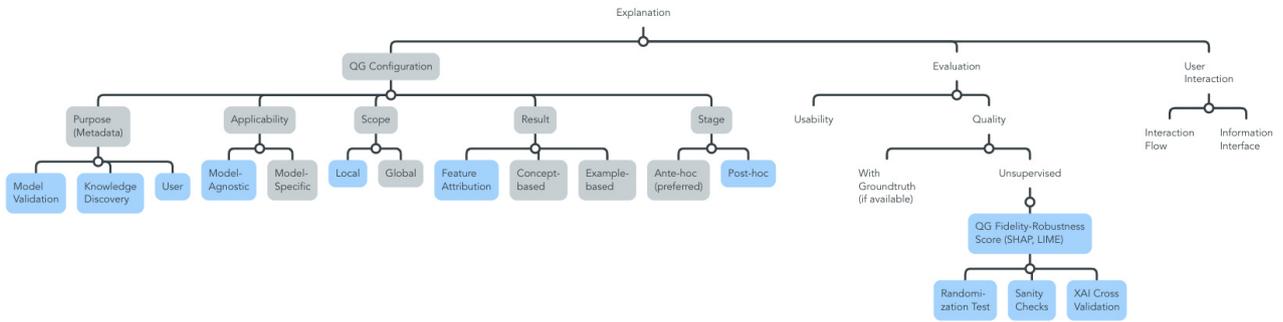

**Figure 6:** Derived QGs along the AI lifecycle for the explanation stage during development are illustrated. We introduce two leaf-QGs, they are colorized and marked by *QG*, mirror concrete design decisions, while collection-QGs reflect generalizable process steps that include sub-levels of collection-QGs and/or leaf-QGs. Leaf-QG *Configuration* provides a general information structure, and leaf-QG *Fidelity-Robustness Score*, a concrete implementation guideline to assess the quality of explanations, as a design knowledge contribution. The general leaf-QG *QG Configuration* has multiple possible realizations. Their selection can be executed according to the IEEE XAI Framework [7]. With respect to the concrete guideline QG *Fidelity-Robustness Score*, the blue nodes are relevant for application. We propose this structure, but it needs to be tested for applicability based on more use cases. The leaf-QG format for information processing is explained in more detail in section 5.

- *Scope* refers to the breadth of the explanation, whether global (covering the entire model) or local (focused on individual predictions). The scope should be selected based on the level of detail needed for the specific audience or task.

- *Result* denotes the form in which the explanation is delivered, such as text, visualizations, or statistical summaries. The result should align with the target audience's needs. For instance, in healthcare, a local explanation could be a visual heatmap showing which parts of an image influenced a diagnosis, providing actionable insights for medical professionals.

- *Stage* describes when and how explanations are generated in relation to model development. It indicates whether explanations are derived from a trained model directly (post-hoc) or require modifications to the model itself (ante-hoc), and affects steps such as evaluation and user interpretation. Typically, post-hoc explanations requiring more intensive validation to ensure their reliability and meaningfulness than ante-hoc intrinsically explainable methods.

These characteristics should be selected based on the specific use-case requirements, such as the model's explainability needs and best practices. They also depend on contextual factors like computational resources, model complexity, and the interpretability needs of stakeholders. Depending on the use case, possibly, a combination of explanations is applied, which results in multiple MQG4A-template versions, if they are tested for the same purpose or multiple leaf-QG configuration within the same MQG4A-version, for different explanation purposes that apply simultaneously.

For example, if the use case requires feature importance (result) and local (scope) explanations, methods like SHAP or LIME can be selected based on these needs. Additionally, both SHAP and LIME explanations can be used for purposes like model validation, model discovery, or user interpretation (purpose), are model-agnostic (applicability), and are post-hoc (stage). These options are highlighted in blue in Figure 6. By considering these factors, MQG4AI's lifecycle structure ensures that the explanation method(s) chosen align(s) with the intended goals and provides the necessary clarity and actionability for the given context.

**Collection-QG *Evaluation*** emphasizes the critical role of assessing explanations in terms of both usability and quality. In contrast to *Configuration*, this QG-type comprises a collection of approaches that produce results to evaluate one XAI method within MQG4A. This setup can be repeated for *n* XAI methods within a single project. MQG4DK simply collects explanation evaluation approaches. This is essential because some explanation techniques, while effective in certain contexts, may fail to meet the specific requirements of a given use case [41]. Therefore, it is important to map these use-case requirements to relevant evaluation metrics to assess the desired qualities of the explanations, ensuring they are suitable for their intended purpose.

However, the evaluation process is not without challenges. It is complex due to domain-specific constraints, the broad range of properties that can be evaluated, and the difficulty in integrating case-specific metrics. Additionally, the lack of generalizable methods for automating this evaluation further complicates the task [4]. Despite these challenges, Collection-QG *Evaluation* provides the foundation for a thorough and effective evaluation process, ensuring that explanations are reliable, aligned with the specific goals and requirements of the use case, and that identified risks are meaningfully mitigated rather than superficially addressed. Evaluation QGs, as summarized in Figure 6, are subdivided int assessing *Usability* and *Quality* of explanations:

- *Usability*, or subjective evaluation, focuses on how understandable and interpretable the explanation is for

the intended audience. It examines factors such as user satisfaction, ease of comprehension, and the extent to which the explanation fosters trust and transparency. These aspects are typically assessed using qualitative methods, including user surveys, case studies, and focus groups. Usability can be further evaluated in terms of parsimony (conciseness), coverage (comprehensiveness), and overlap (clarity) [28].

- *Quality*, or objective evaluation, focuses on the accuracy and reliability of the explanation itself. As defined in Figure 6, it can be measured using the explanation's ground truth (if available), or in an unsupervised manner. In the absence of the explanation's ground truth, mathematical properties can be identified and calculated through computational techniques, providing a systematic and quantitative evaluation of explanation reliability. These properties act as building blocks that form a robust framework for evaluating explanations. For instance, [30] identifies two key and interrelated explanation properties (robustness and fidelity) and proposes a set of computational methods to systematically evaluate SHAP and LIME explanations on both dimensions.

Finally, **User-Interaction Collection-QGs** enforces that explanations are effectively presented and clearly communicated to relevant stakeholders, elucidating the intricacies of the decision-making process behind the specific decisions they reference. This stage focuses on how to present the generated explanations, which builds on user studies and usability testing to gather actionable insights into user interactions in alignment with explanation usability evaluation. The overall objective comprises well-designed user interfaces (UIs).

To operationalize this *Explanation* lifecycle we can instantiate leaf-QGs. For example, to assess the quality of LIME and SHAP explanations, we propose the leaf-QG *QG_FidelityRobustnessScore_(SHAPLIME)*. This concrete design decision is inspired by [30] and outlines how to implement unsupervised methods within MQG4DK to evaluate metrics like fidelity and robustness, resolving the robustness problem associated to explainability techniques. It will be retrieved within dynamic MQG4A-versions if LIME and/or SHAP explanations are applied, and we use it as an example to illustrate how to define leaf-QGs that mirror concrete lifecycle design decisions based on existing RAI knowledge in form of a published technical guideline in section 5.

After introducing MQG4AI's *Explanation* lifecycle stage that incorporates an application planning-oriented view how to integrate expainability-related processes and design decisions with the generic AI lifecycle in form of general collection- and leaf-QGs, while incorporating best practices towards risk mitigation, we evaluate our proposition in the next section from a conceptual viewpoint.

### 4.3. Evaluation of the Proposed *Explanation* Lifecycle Information Structure

We evaluate the proposed *Explanation* lifecycle stage based on the *IEEE Guide for an Architectural Framework for Explainable Artificial Intelligence* (XAI) [7]. Overall, our approach aligns with the IEEE guidelines and offers comprehensive mechanisms to address all relevant processes for achieving explainability, as detailed in this section.

**IEEE XAI Framework** The guide outlines a comprehensive compilation of methods that define the *Explanation* lifecycle and related stages. While they aim for overall AI system transparency, they focus on "[...] adopting a variety of XAI methodologies" [7, 2], highlighting a reliable XAI method definition in alignment with the intended use, as well as means for XAI method evaluation. According to the framework explainability is influenced by three aspects: Content type, i.e. what to explain, Communication, i.e. how to explain, and Stakeholders, i.e. to whom the explanation is addressed [7, 14]. By addressing all three dimensions, the framework outlines when and how different stakeholders should engage with explanations across the three stages of the AI lifecycle: pre-modeling, modeling, and post-modeling. During pre-modeling, developers focus on XAI interdependencies with data quality, and quality assessment of the underlying data that impacts the explanation's quality. Modeling refers to explanations that can happen in parallel to model development, a decision made by the developer.[13] Post-modeling summarizes steps to explain the model after it is trained, which comprises additional stakeholders such as the customer and subject[14] to evaluate the choice of explanation. [7, 20] Finally, explanations are assessed in terms of qualitative, quantitative, subjective, and objective evaluation metrics. The first two are based on user usability and experience, while the second two consider explanation quality and can be formulated by mathematical language. [7, 23]

We identified the following correspondence of the proposed MQG4AI information structure with knowledge provided by the IEEE XAI framework regarding (1) the three identified aspects, (2) the modeling stages, as well as (3) methods for explanation assessment. Overall, we aim to provide a structure that aligns lifecycle processes and design decisions with supplementary information to continuously plan a reliable implementation.

**Content, Communication, Stakeholder**: The three aspects Content, Communication, and Stakeholder [7, 14] are referenced through leaf-QG configuration. Content (what) corresponds to *purpose*, and Communication (how) comprises the other criteria (*applicability*, *scope*, *result*, *stage*) that conclude the explanation configuration, as previously introduced, and illustrated in Figure 6. Stakeholder (to whom) is implicitly derived based on the identified *purpose* of the explanation configuration, which is further clarified within the individual leaf-QG template, as explained in the next section 5, and depicted in Figure 7. Also, QG user interaction contributes to Communication and is adjusted to Stakeholder that are not developer. In addition, to enable information linking, MQG4AI summarizes contextual information on stakeholders and TAI within the supplementary information block sections system information and risk management, as introduced in section 3. This information is equally referenced within the individual leaf-QGs.

---

[13]Ante-hoc methods are equally an option for DNNs.

[14]AI subjects comprise "[...] organizations or entities who are impacted by the AI products or services [...]" [7, 21].

**Modeling-stages**: MQG4AI implements the defined pre-modeling, modeling, and post-modeling stages of the AI lifecycle through collection-QGs (MQG4DK & MQG4A) and successive template versions (MQG4A). Pre-modeling refers to data quality and preprocessing steps during the data utilization stage, see Figure 5. Information on horizontal interdependencies between lifecycle design decisions is included within the leaf-QG template's *interdependency graph*, as depicted in Figure 7, adding information extraction for other lifecycle stages, and beyond the development stage. Modeling refers to data modeling (model generation), which references information on the other three model development stages (configuration, evaluation, optimization), as depicted in Figure 5. They equally comprise horizontal interdependencies and are linked with the data utilization stage. This intricacy is illustrated in Figure 4. Finally, post-modeling, as proposed by IEEE focuses on post-hoc methods regarding explanation generation by the AI developer and evaluation by the subject and customer. The latter is included with the explanation lifecycle collection-QG user interface and usability evaluation.

Note that the modeling and post-modeling stages distinguish between the generation of ante-hoc and post-hoc explainability, as proposed by the IEEE framework. In contrast, the MQG4AI-template integrates both explainability types under the unified leaf-QG configuration for lifecycle planning, merging them into the post-modeling stage while separating them from the modeling stage. This integration reflects our perspective that model development and explanation implementation are treated as distinct processes, allowing for parallelization within the AI lifecycle. By decoupling these stages, we facilitate a more modular and efficient workflow. Further, intrinsically explainable models can equally undergo evaluation for usability or quality, independently of their role in the broader lifecycle. However, in the case of an ante-hoc approach, model development and explanation implementation are linked more closely. Some authors even argue that, as of now, "[m]ost AI algorithms have not been designed (or evolved) with intrinsic methods, therefore XAI is primarily a collection of extrinsic methods [...]" [8, 8]. In support if this design choice, the lifecycle planning template is consulted in a comprehensive manner as part of iterative lifecycle conceptualization stages, which results in a non-linear execution of all stages, since they are organized together with different focuses depending on the overall project planning status, as explained in section 3.

**Explanation evaluation**: Regarding explanation evaluation, the reviewed IEEE framework considers qualitative, subjective, quantitative and objective evaluation metrics, which can be directly aligned with our explainability evaluation collection-QG (usability and quality). Subjective evaluation is related to usability, as it involves assessing whether the results satisfy the user's requirements. In addition, IEEE includes user assessment regarding regulatory compliance based on individual knowledge of the user on their rights [7, 25], which we do not clarify in more detail.

Moreover, objective evaluation is related to quality, as it includes a set of mathematical properties that are calculated with computational techniques. Regarding our proposed design decision for explanation evaluation in the next section 5, fidelity is explicitly included within the IEEE framework, whereas what we call robustness can be interpreted as a concrete implementation to calculate sensitivity, measuring "[...] the degree to which the interpretation is affected by the small disturbance from the test point" [7, 12].

Further, IEEE contributes comprehensive knowledge on model-specific methods that are intended to provide insights into a model's decision-making process tailored to its architecture. In addition, they outline characteristics for data-centric methods that focus on improving the quality of training data to enhance model performance. This knowledge can function as foundation to further populate the MQG4AI-template structure, as well as how to fill it for MQG4A-scenarios. For instance, gradient-based model-specific methods are used for DNNs and belong to applicability - model-specific within leaf-QG configuration, as depicted in Figure 6.

Additional evaluation assesses transparency across several dimensions. These include simulatability, which examines how the model's behavior changes with varying data [7, 23]; decomposability, which focuses on understanding the individual components of the model; and algorithmic transparency, which ensures clarity on how specific input combinations produce particular outputs [7, 24]. These aspects collectively enhance the interpretability and trustworthiness of the model, but are not directly related to explainability. Generally, the MQG4AI-template is envisioned to contribute information to all of these transparency-related aspects through comprehensive lifecycle planning, versioning and interdependent information processing. The explanation lifecycle stage focuses on the application of concrete methods for design decision-making with a specific purpose, contributing to transparency.

After introducing and evaluating MQG4AI's explanation stage, we demonstrate the leaf-QG template in the next section. Namely, we outline how to contribute knowledge for explanation assessment to MQG4DK.

## 5. Leaf-QG Fidelity-Robustness-Score to Evaluate the Quality of Explanations

The aim of this section is to introduce the leaf-QG format, as demonstrated in Figure 7, with the proposed leaf-QG *QG_FidelityRobustnessScore_(SHAPLIME)* as an example. It is equally included in GitHub as part of the MQG4AI-template. According to the *QG-Naming-structure*, this QG contributes a Fidelity-Robustness-Score to evaluate SHAP and LIME generated explanations to the generalizable requirement to evaluate the quality of explanations. This *Explanation* lifecycle stage collection-QG addresses the risk *Unfaithful Explanations* which contributes to overall *Transparency*. Other contributions to collection-QG *XAI Quality Evaluation* can append the measurement of different methods, or extend the existing Fidelity-Robustness-Score with further information for other XAI-methods, for instance.

| QG Name_(View) and Tags | | | |
|---|---|---|---|
| Interdependency Graph | Input Information | Related Lifecycle Implementation QGs | |
| | | AI System Information (Application, Stakeholder, ...) | |
| | Output Information | Related Lifecycle Implementation QGs | |
| | | Post-Market Monitoring System Information | |
| QG Creation (Dimensions) | | Content (Which information is generated?) | |
| | | Method (How is the information generated?) | |
| | | Representation (How should the information be presented?) | |
| | | Stakeholder 1 | ... | Stakeholder n |
| | | Evaluation (Are there open questions?) | |
| Additional Information | Risk Management | Poses Risk | |
| | | Implements Risk Control | |
| | ... | | |

**Figure 7:** Proposed leaf-QG template for RAI design decision information processing.

The **Interdependency Graph** includes related *Input and Output Information* that is extracted from other lifecycle stages, and supplementary sections, focusing on interdependencies. **Input Information** answers the question: *What information is necessary to execute the method and generate the content?* The defined QG requires SHAP or LIME explanations that are derived from predictions of a previously trained model. Intuitively, the underlying model needs to allow the generation of such explanations, which in some cases can pose some restrictions for the model choice. This could also serve as input information. If analyzed in further detail, the selected model and data, and its subsequent preprocessing and performance, affect the quality of the input explanation. For this reason, related lifecycle implementation QGs, as depicted in Figure 5, include:

- *QG_Utilization_(Data)* comprises information on data quality through statistical analysis, as well as preprocessing steps.

- *QG_Configuration_(Development)* provides knowledge on the model whose AI output is explained.

- *QG_Evaluation_(Development)* comprises performance metrics that are used to monitor the model, which impacts the quality of explanations, as illustrated in Figure 4.

- *QG_MethodConfiguration_(Explanation)* defines the implemented explainability technique that is evaluated.

In addition, contextual *AI System Information* can be linked. However, in this case, we did not identify any AI system-specific information.

**Output Information** responds to the question: *Which stages are impacted and which additional information might be necessary?*, focusing on extracting relevant information for post-market monitoring. The proposed QG assesses explanation quality by integrating fidelity and robustness as evaluation criteria. As noted by [30], explanations lacking fidelity risk misleading users by creating a false sense of understanding. To address this concern, the application-oriented contribution introduces a unified quality score that evaluates fidelity and robustness simultaneously, ensuring a more thorough assessment. As mentioned, the proposed QG evaluates explanation quality in terms of fidelity and robustness. The output information for post-market monitoring answers, on a scale from 0 to 1, the question: *Are LIME/SHAP explanations appropriate for explaining the model?*, which is triggered, as soon as the model is retrained on new data, and thus, needs to be included within the monitoring strategy. A score of 0 indicates that LIME/SHAP explanations are not appropriate for the task, and should not be trusted, whereas a score of 1 indicates that LIME/SHAP explanations are perfect for the task, and any score in between can suggest that explanations are either appropriate or not. For the last cases, a suggested rule of thumb is that everything above 0.8 is good, and above 0.9 is pretty good. If the score is lower, it is advisable to regenerate the explanations for the ML task, possibly altering the model, data, or feature importance method used. Consequently, impacted generic lifecycle collection-QGs comprise:[15]

---
[15]*Deployment* and *Maintenance* are not focus of our contribution, and therefore only referenced. With advancing knowledge, more precise

- *QG_MethodConfiguration_(Explanation)* may need adjustment if the generated explanations do not meet the required standards, based on metric results.

- *QG_Deployment* requires validation of the quality and accuracy of explanations, linked to post-market monitoring data to assess real-world performance.

- *QG_Maintenance* focuses on continuous monitoring and evaluation, ensuring improvements are made based on post-market insights.

The **QG Creation** comprises three information dimensions, including an *Evaluation* section to assess the chosen method: *Content* explains the design decision in more detail, *Method* outlines design decision-making process, and *Representation*, which focuses on transforming information for relevant stakeholders. Overall, this proposed information processing from different perspectives is intended to contribute towards quality by design, since it propagates rethinking individual design decisions.

1. *Content*: To evaluate fidelity and robustness, a set of sanity checks and robustness measures to distribution changes are employed, respectively. Here, any assumptions made by the method should be specified. In this case, the method assumes that if explanations are robust to different test sets, they are also robust to distribution changes in the test set (as the only way the distribution of the test set can change is if the test set changes). Moreover, if explanations have passed the fidelity checks, then the mean of the generated explanations is representative of the explanations. In the proposed QG, fidelity evaluation outputs a score of either 0 or 1, and robustness evaluation a value between 0 and 1, both being afterwards combined by the multiplication operation.

2. *Method* In this dimension, further details on how to implement the proposed QG are included. **Sanity checks** are implemented by a set of randomization tests (model randomization and data randomization checks) that assess whether the explanations are genuinely informative about the model and data [2]. The **data randomization test** compares explanations generated from a model trained on original data (base scenario) against those generated from a model trained on data with randomized class labels. The **model parameter randomization test** compares explanations generated under the base scenario against those generated from a randomly initialized model. For both cases, feature importance explanations can be considered similar if their feature importance rank is similar. However, since less important features contribute less to the model's overall behavior it is especially important to ensure that the feature importance rank is similar for the most important features. According to [30], the proposed QG will use the **Normalized Discounted Cumulative Gain (NDCG)** metric to quantify the similarity between explanations (as it prioritizes the correct ranking of higher-importance features). Additionally, **robustness** is measured by calculating the similarity of explanations that are generated from data of slightly different distribution (i.e. trained and tested on slightly different data splits, where each split introduces small variations in the data distribution). Similarly, the NDCG metric will be used to quantify the similarity between these explanations. Finally, both results will be merged through multiplication to obtain a final score, which would be 0 if fidelity is 0, and otherwise (if model fidelity exists) would mirror the explanation's robustness to changes in the data distribution.

3. *Representation* The last dimension addresses which information should be presented to which stakeholders and when. Identified stakeholders include AI Experts and Data Scientists (active), to reconsider data preprocessing, model development or explanation generation; Domain experts (consulting), to validate explanations subjectively; Regulators (passive), to assert the compliance of the AI system with current regulation; and AI Users (passive), to interprets model decisions and understand underlying decision-making process.

4. *Evaluation* comprises open questions regarding the design decision, which can function as foundation for residual risk evaluation, for instance. Fidelity and robustness metrics were chosen because they are frequently referenced in the literature and can be applied to evaluate various types of explanations. Further, the proposed method is independent of domain and use case. The method's underlying assumption is that SHAP and LIME generate feature importance explanations, which highlight the significance each feature holds in model predictions. Therefore, the focus lies on the similarity of feature importance ranks, as less important features are likely less informative about the model. Consequently, we define the similarity of explanations (NDCG score) in terms of both the similarity of feature ranks (order of features by importance value) and the feature importance values, ensuring it remains unaffected by the specific topic or use case. For this example, SVM, MLP, and XGBoost models are used, and the approach is tested across two different datasets. [30]

Finally, **Additional Information** includes a *Risk Management* section for information linking. The related risk *Unfaithful Explanations* is assigned to the TAI criteria *Transparency* and the related subsection *Explainability*, following

---

process steps can be identified. This approach equally mirrors different stages of MQG4A-template conceptualization versions with advancing focus towards the model's real world integration, the following stages will be defined in more detail, and interdependencies with model development refined.

the TAI structure, as proposed by ALTAI [22]. Overall, the chosen fidelity and robustness metrics are thought to limit risks in a reliable manner. This is thanks to their frequent reference in literature, the core and complementary qualities of explanations that they measure, and their flexibility for different types of explanations. In addition, the score provides an application-oriented approach, and considering these criteria, the approach is considered a reasonable contribution to MQG4DK. Nevertheless, the question of whether they are the optimal way to evaluate explanation remains, although as demonstrated by [30], they can effectively compare the quality of feature importance explanations in a domain- and use-case-independent manner.

Concluding, this section illustrated a contribution to MQG4DK based on a concrete design decision to the previously in section 4 introduced generic *Explanation* lifecycle stage that incorporates best practices by design for responsible AI lifecycle planning. In combination with information linking of related AI risk management, this is how we propose a possible workflow for MQG4AI-template design and interaction. The next section discusses our contribution focusing on how the MQG4AI-template corresponds to AI Act-conform quality management systems through the introduced information management structure that incorporates RAI knowledge.

## 6. Discussion

While the current template version focuses on AI RM, we envision the proposed structure to be extendable regarding other AI QMS layers based on the proposed IM structure. In addition, we aim to promote the inclusion (MQG4DK), and implementation (MQG4A) of comprehensive methods that adhere to characteristics of AI Act-aligned standards, as introduced in [38]. For instance, the leaf-QG template information structure promotes a "[s]ufficiently prescriptive and clear" [38, 3] design decision, and MQG4AI covers the complete AI lifecycle, while providing the option to incorporate AI Act requirements through information management. The overall setup, focusing on the identification of interdependencies for information docking, equally aims to facilitate the integration of AI-related information with existing QM processes outside the MQG4AI template. They possibly already organize relevant data, regarding QM adhering to sector-specific regulations, for instance. As a starting point, we focus on linking design decisions with AI risks, which we illustrate organized according to TAI [22] and system-specific information based on the AIRO [17].

Therefore, MQG4AI already incorporates AI QMS requirements that are outlined in Article 17 of the AI Act [13], and we are confident that MQG4AI can be adapted to contribute to all related regulatory requirements through lifecycle planning through comprehensive IM:

- The inclusion of RM is central to MQG4AI, as previously discussed. The necessity for continuous updates and iterations of identified RM processes and risks is further explored in alignment with reflecting the iterative nature of lifecycle design decision-making, and information linking.

- Regarding *Data Management*, the data lifecycle is a component of the proposed lifecycle within the generic MQG4AI-concept, which needs to be further outlined.

- Further, information contributions are extracted during development that contribute to the *Post-market Monitoring System (PMMS) (Article 72)* through information extraction and linking, incorporating the technology's inherent dynamics.

- MQG4AI focuses on the interface between regulation and development. It is designed to support the four requirements (*Strategy for Regulatory Compliance; Design, Design Control, Design Verification; Development, Quality Control and Quality Assurance; Continuous Examination, Test and Validation procedures*), which needs to be tested and evaluated through practical high-risk AI projects utilizing the template for lifecycle planning as part of future work.

- Finally, based on MQG4AI's customizability, highlighting additional leaf-QG layers, as well as appending information sections, the final six requirements (*Technical Specifications; Reporting of serious incidents (Article 73); Communication with National Competent Authorities; Record Keeping; Resource Management; Accountability Framework*) can be addressed analogously.

## 7. Conclusion

In this paper, we introduce the proposed generic and customizable MQG4AI-template structure for continuous AI lifecycle conceptualization, available on GitHub.[16] Our contribution functions as a starting point, aiming to illustrate relating design decision-making with other relevant high-risk RAI information towards overall risk mitigation by design. In the present contribution, we explore reliable XAI-evaluation metrics, and attempt to propose a generalizable process structure of the *Explanation* stage including vertical and horizontal interdependencies through the identification of collection- and leaf-QGs that construct the AI lifecycle. In addition, we demonstrate the leaf-QG information structure for a concrete design decision centered around evaluating the quality of explanations.

Overall, MQG4AI's decentralized, and adaptable character is intended to respond to AI's novelty, inherent dynamics, and use case-specificity, so that methods for implementation can be structured, and continuously stay up-to-date – including future contributions. This is achieved through MQG4AI's DSR setup, its structured information flow including contextual information, different template versions, the QG-format for information organization and processing along the AI lifecycle, as well as customizability of MQG4AI's building blocks. We envision to enable the creation of a growing RAI ontology to organize generalizable instructions for the domain- and use case-adapted implementation in form of MQG4DK. In parallel, MQG4A is intended to provide a practical, and generic approach that enables stakeholders to oversee AI lifecycle implementations in a responsible manner through the incorporation of ethics and regulatory requirements towards quality by design for high-risk systems.

---
[16] https://github.com/miriamelia/MQG4AI

Our goal is to assess whether MQG4AI is perceived as a valuable approach. Moving forward, refining the basic structure to include additional risks, use cases, and lifecycle stages will require contributions from a more diverse range of experts. Additionally, the current implementation on GitHub needs further development towards becoming a software that can be utilized. For instance, automating bidirectional information linking and creating an intelligent search feature to retrieve configurable MQG4AI-versions based on leaf QG-tags are essential improvements. Future work should focus on refining all proposed sections, testing the concept through a concrete (medical) AI project, and developing a software that facilitates MQG4DK and MQG4A interplay, along with a user-friendly, multi-stakeholder interface.

## A. Declaration of generative AI and AI-assisted technologies in the writing process

During the preparation of this work the author(s) used *ChatGPT* in order to to improve the readability, grammar and language of the manuscript. After using this tool, the authors reviewed and edited the content as needed and take full responsibility for the content of the published article.

## CRediT authorship contribution statement

**Miriam Elia:** Conceptualization, Writing – original draft, Investigation, Formal analysis, Methodology, Writing – review and editing, Software. **Alba Maria Lopez:** Software, Formal analysis, Investigation, Writing – original draft. **Katherin Alexandra Corredor Páez:** Methodology, Formal analysis, Software, Writing – review and editing. **Bernhard Bauer:** Conceptualization, Funding acquisition, Supervision, Writing – review and editing. **Esteban Garcia-Cuesta:** Conceptualization, Supervision, Funding acquisition, Writing – original draft, Writing – review and editing, Investigation.